\documentclass[12pt,preprint]{aastex}

\shorttitle{FUV observations of the globular cluster NGC\,2808}
\shortauthors{Dieball et al.}

\begin{document}

\title{Far-Ultraviolet Observations of the Globular Cluster 
NGC\,2808 Revisited: Blue Stragglers, White Dwarfs and
Cataclysmic Variables \footnote{Based on observations made with the NASA/ESA 
{\it Hubble Space Telescope}, obtained at the Space Telescope Science
Institute, which is operated by the Association of Universities for
Research in Astronomy, Inc., under NASA contract NAS 5-26555.}}

\author{A. Dieball, C. Knigge} 
\affil{Department of Physics and Astronomy, University of Southampton,
  SO17 1BJ, UK} 

\author{D. R. Zurek, M. M. Shara}
\affil{Department of Astrophysics, American Museum of Natural History,
New York, NY 10024}

\and

\author{K. S. Long}
\affil{Space Telescope Science Institute, Baltimore, MD 21218}

\begin{abstract}

We present a reanalysis of far-ultraviolet ($FUV$) observations of the
globular cluster NGC\,2808 obtained with the {\it Hubble Space
  Telescope}. These data were first analyzed by Brown and coworkers,
with an emphasis on the bright, blue horizontal branch (HB) stars in this
cluster. Here, our focus is on the population of fainter $FUV$
sources, which include white dwarfs (WDs), blue stragglers (BSs) and
cataclysmic variables (CVs). We have therefore constructed the deepest
$FUV-NUV$ colour-magnitude diagram of NGC\,2808 and also searched for
variability among our $FUV$ sources. Overall, we have found $\approx 40$ WD,
$\approx 60$ BS and $\approx 60$ CV candidates; three of the BSs and
two of the CV candidates are variable. We have also recovered a known
RR Lyrae star in the core of NGC\,2808, which exhibits massive
($\approx$~4 mag) $FUV$ variability. We have investigated the radial
distribution and found that our CV and BS candidates are more
centrally concentrated than the HBs and WD candidates. This might
be an effect of mass segregation, but could as well be due to the preferential
formation of such dynamically-formed objects in the dense cluster
core. For one of our CV candidates we found a counterpart in
WFPC2 optical data published by Piotto and coworkers.  

\end{abstract} 


\section{Introduction}
\label{intro}

Globular clusters (GCs) are old, gravitationally bound stellar
systems whose core stellar densities can be extremely high,
reaching up to $10^{6} {\rm stars}/{\rm pc}^{3}$. In such an
environment, close encounters and even direct collisions with 
resulting mergers between the cluster stars are quite common, leading
to a variety of dynamically-formed stellar objects like Blue
Stragglers (BSs) and close binary (CB) systems. CBs are important for
our understanding of GC evolution, since the binding energy of a few,
very close binaries can rival that of a modest-sized globular
cluster. Thus, by transferring their orbital energy to passing single
stars, CBs can significantly affect the dynamical evolution of the
cluster (e.g.\ Elson et al.\ 1987, Hut et al.\ 1992, and references
therein). This depends critically on the number of CBs. If there are
only a few CBs, long-term interactions dominate the cluster
evolution. By contrast, the presence of many CBs leads to violent
interactions, which heat the cluster, and promote its expansion and
evaporation.  

Interacting CBs form a particularly interesting subset of CBs, and
can, in principle, also be used as tracers of a cluster's CB
population. The best-known formation channel for
interacting CBs in GCs is tidal capture of a red giant or main
sequence (MS) star by a compact 
object \citep{fabian}. This scenario was originally proposed to
account for the overabundance of low-mass X-ray 
binaries (LMXBs) containing accreting neutron stars (NSs) in GCs,
relative to the galactic field. However, tidal capture theory
predicts a comparable overabundance of interacting CBs with a white
dwarf (WD) primary, i.e. cataclysmic variables (CVs). Since WDs are
far more common than NSs, we would then also expect many more CVs than
LMXBs in GCs. More recently, it has been realized that 3-body encounters
(e.g. Davies \& Benz 1995) and ``ordinary'' evolution of  
primordial binaries (e.g. Davies 1997) may also produce significant
populations of CVs in GCs. CVs produced from primordial binaries are
expected mainly in the outskirts of clusters, whereas CVs formed
dynamically (through tidal capture or 3-body processes) should be
found preferentially in the dense cluster cores. Thus, the observation
of the relative abundances and distribution of CVs within a cluster's
core and halo can tell us about the relative efficiency of these
different CV formation scenarios. 

However, despite their impact on cluster evolution and their
importance for our understanding of CB formation and evolution
channels, there have been only a few detections of interacting CBs in 
GCs during the past decades. Finding and studying these systems has
proven to be extremely difficult, since the spatial resolutions and
detection limits of most available telescopes are too limited for their
detection. Only with the improved sensitivity and imaging quality of
{\it Chandra} and {\it HST} has it become possible to finally detect
significant numbers of CVs and other binary systems in GCs (e.g.\
Grindlay et al.\ 2001; Albrow et al.\ 2001; Edmonds et al.\ 2003a,
2003b; Knigge et al.\ 2002, 2003; and references therein).  

$FUV$-imaging is particularly well-suited for the detection of CVs (and
also BSs and young WDs) in GC cores. This is because these objects
are characterized by relatively blue spectral energy distributions and
emit significant amounts of radiation in the $FUV$. By contrast, the
``ordinary'' stars that make up the bulk of the cluster are too cool
to show up in observations at such short wavelengths. As a
consequence, crowding is not a severe problem in $FUV$ imaging studies
of GCs, even in the dense cluster cores. This is a significant
advantage compared to optical GC surveys.

To date, the only deep $FUV$ study of a globular cluster with the
principal purpose of identifying CVs (as well as BSs and WDs) has been
of 47~Tuc \citep{knigge1,knigge2}. There, we found 16 CV candidates
(including 4 variable objects that were previously known or suspected
cataclysmics), 19 BSs (including 4 variables) and 17 hot WDs. The
population of CV candidates was particularly interesting, since their
number was broadly consistent with tidal capture predictions (but note
that most of these candidates have not yet been confirmed).  

Here, we present a reanalysis of {\it HST}-based $FUV$ observations of another
GC, namely NGC\,2808 ($\alpha = 09^{h} 12^{m} 02^{s}$, $\delta = -64^{\circ}
51{\arcmin} 47^{\arcsec}$). This intermediate metallicity GC 
($\rm{[Fe/H]} = -1.36$, Walker 1999) lies at a distance of
10.2 kpc and is reddened by $E_{B-V}=0.18\pm0.01$ mag (Bedin et
al. 2000). The cluster has a very dense and compact core, with a core 
radius of $0\farcm26$ and a tidal radius of $15\farcm55$
\citep{harris}. 

The data set that we analyse in the following has already been studied by 
\citet{brown}, but with a focus on the bright, blue HB stars in the
cluster core. More specifically, \citet{brown} discovered a population
of subluminous horizontal branch stars in NGC\,2808 which might have
undergone a late Helium flash while descending the WD cooling curve. 
By contrast, here we are mainly interested in the dynamically-formed
stellar populations (CVs and BSs), as well as the young WDs in this
cluster. Since these are faint objects and normally difficult to
detect, we first construct a new, deep colour-magnitude diagram (CMD)
in which many of these objects show up, and then search for
variability amongst our $FUV$ sources (since this is ubiquitous amongst
CVs and also seen in BSs located in the instability strip).

In Sect.\ \ref{data}, we discuss the data and the data reduction. In
Sect.\ \ref{results}, we present our far- vs. near-ultraviolet CMD and
describe our search for variability among our catalogue stars. Our
results are summarized in Sect.\ \ref{summary}.  

\section{Observations and data reduction}
\label{data}

\begin{figure}
\plottwo{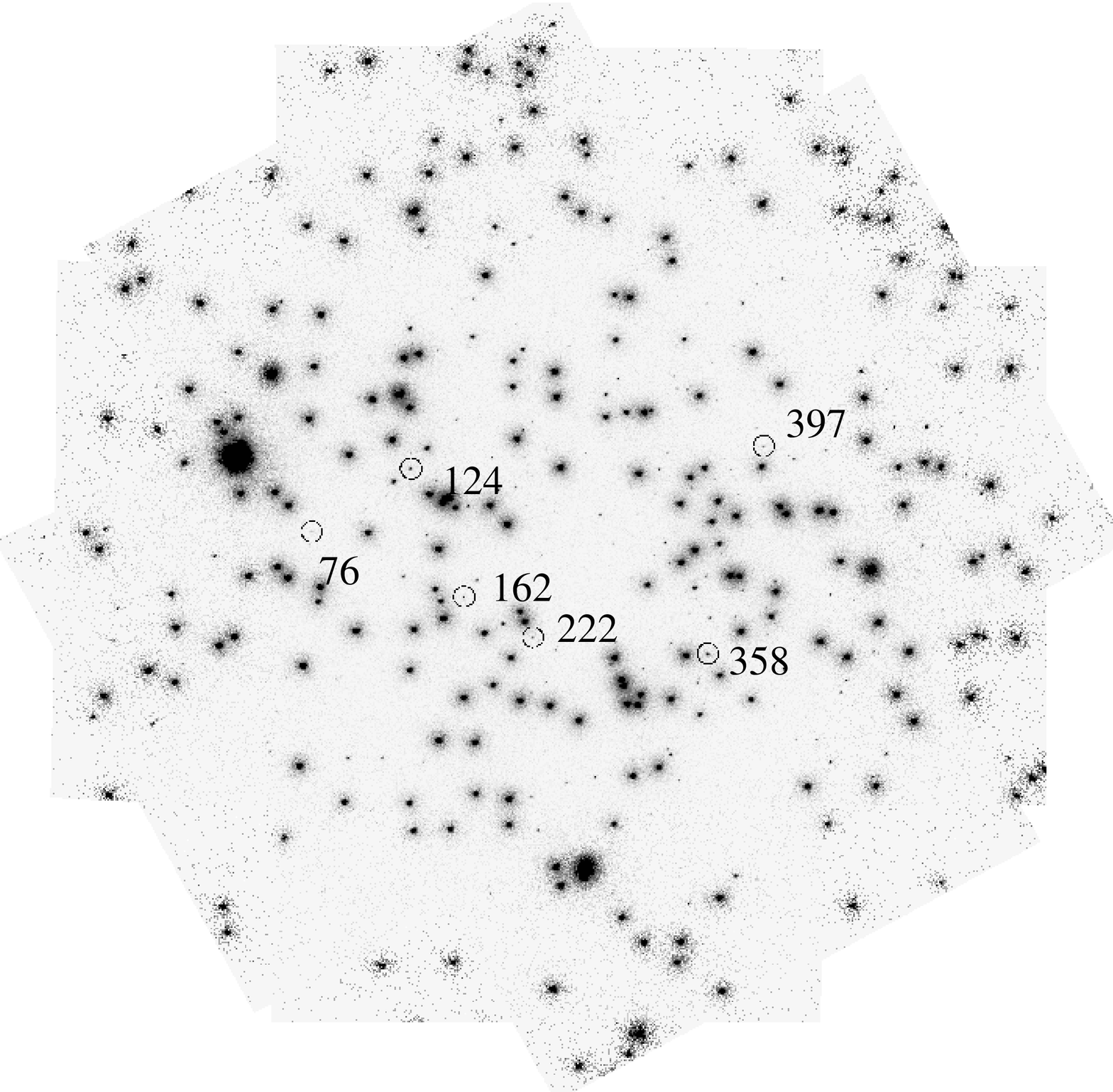}{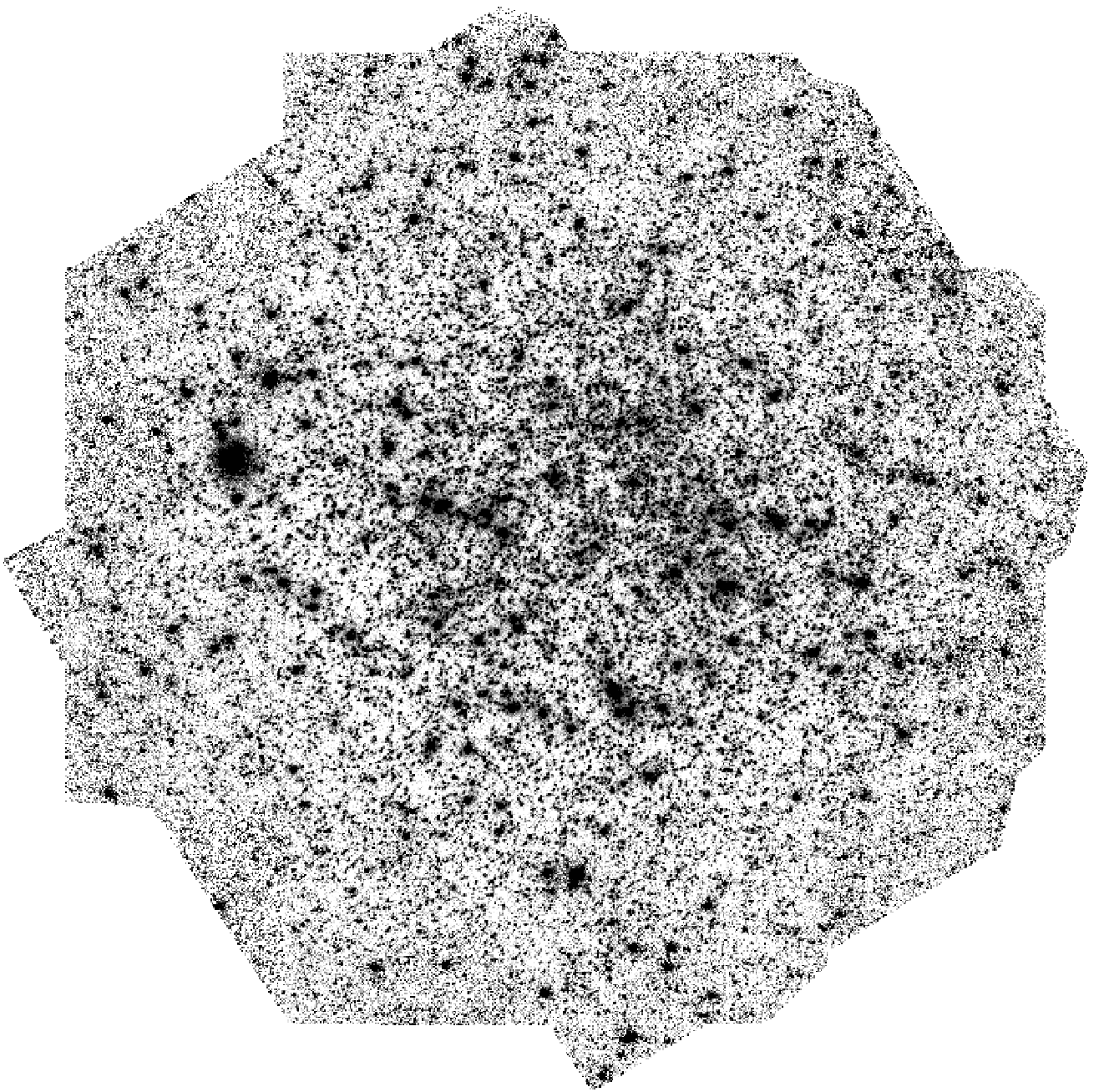}
\caption{\label{fuv} Left: Mosaic of all $FUV$ exposures taken from
  NGC\,2808's core. The single exposures were taken with slight offsets and
  rotation with respect to each other which led to the strange shape
  of the mosaic. Variable candidates are marked with black circles and
  their corresponding catalogue number in Table~\ref{catalogue}; see also
  Sect.~\ref{variability}. The images are displayed on a logarithmic
  intensity scale in order to bring out the fainter sources and
  illustrate the non-uniform background caused by the varying exposure
  times across the images. Right: Mosaic of all $NUV$ exposures of the
  same region of NGC\,2808. Note the severe crowding in the $NUV$.}   
\end{figure}

NGC\,2808 was observed with the Space Telescope Imaging
Spectrograph (STIS) on board the {\it HST} in
January/February 2000. The original purpose of these observations was 
to provide geometric distortion corrections for the STIS
camera. However, such calibration data are of course also of
scientific interest, as already shown by \citet{brown}. 
Images were taken using the F25QTZ 
and F25CN270 filters, which are located in the far- and near-UV
wavebands, with central wavelengths of 1590 \AA\ ($FUV$, F25QTZ) and
2700 \AA\ ($NUV$, F25CN270), respectively. The plate scale is
$0\farcs0248/\rm{pixel}$, resulting in a field of view of $25\arcsec
\times 25 \arcsec$ for both filters. The exposure times for the single
exposures vary between 480 and 538 sec. The observations were positioned on
the cluster centre; Fig.~\ref{fuv} shows the combined images. The
mosaic images were constructed using MONTAGE\,2 \citep{stetson94},
which registers all the image frames to a single reference frame -
incorporating shift, scale, rotation, and distortions determined from
positions of the brightest stars.

The resulting $FUV$ and $NUV$ mosaic images show a strange shape that is due
to the small offsets and rotation of the single exposures with respect
to each other. The diameter of the mosaics is $\approx 2100~\rm{pixel}$
or $\approx 52\arcsec$. Since the cluster's core radius is
$16\arcsec$, the complete core is covered by our mosaic images. 

As can be seen immediately, the $NUV$ mosaic is extremely
crowded. Crowding, on the other hand, is not a problem in the $FUV$ 
image. Note that the effective exposure times vary strongly across
both mosaics. This is responsible for the increased background noise
in the outskirts of the mosaics, where exposure times are low compared
to the central regions (which have maximum exposure times of 8906 sec
[$NUV$/F25CN270] and 8916 sec in the [$FUV$/F25QTZ]). 

\subsection{Source detection}
\label{find}

As already mentioned in Sect.~\ref{intro}, we reinvestigate the same
dataset studied by \citet{brown}, but with a different focus. Here,
our main goal is to search for faint and/or variable $FUV$ sources (CVs,
BSs, WDs). We therefore take special care to detect even weak sources
in the $FUV$ images, which allows us to construct a very deep CMD. 

As noted above, the exposure time is not uniform across the
$FUV$ mosaic, which means that the outer parts of the image (with the
lowest exposure times) have the noisiest background. In order to
avoid both a large number of false detections in these regions (caused 
by noise peaks), and multiple detections in the halos of bright stars, 
we created a noise map which we used to ``smooth'' the
image. As a first step, we then used the {\tt daofind} routine of {\sc daophot}
\citep{stetson} running under {\sc iraf} \footnote{{\sc iraf} (Image
  Reduction and Analysis Facility) is distributed by the National
  Astronomy and Optical Observatories, which are operated by AURA,
  Inc., under cooperative agreement with the National Science
  Foundation.} 
on the smoothed image to detect $FUV$ sources.\footnote{Note that the
  smoothed image was only used for source detection; all photometry
  was performed on the unsmoothed images.}  
We overplotted the resulting source coordinates on the  
$FUV$ mosaic and inspected them by eye. Overall, this source finding
technique appeared to work well for bright sources (except for a few
multiple detections around bright stars that we deleted by
hand). However, it did not work as well for faint sources, especially
in the deep, central part of the mosaic. Thus, we added a further 211 
sources by hand. In total, our catalogue of $FUV$ sources contains 521
entries. 

Source detection in the $NUV$ mosaic is a tedious task due to the severe
crowding and depends critically on the {\tt daofind} detection threshold. 
If the detection threshold is too high, no faint
sources will be detected in the centre of the mosaic (which has
maximum exposure time). The situation can be improved by lowering the detection
threshold, but this also leads to false ``detections'' in the noisy
outskirts. (The same is true for the $FUV$ mosaic, but less of a
problem, since crowding is much less severe. We could therefore simply
adopt a relatively high threshold in {\tt daofind} and add the missing
sources by hand.) However, it is important to realize that our goal is
not to detect all $NUV$ sources, but to locate $NUV$ counterparts to
our $FUV$ sources. In practice, we therefore chose to work with an
intermediate threshold, designed to yield most of the detectable
$NUV$ counterparts while keeping the number of false detections at
bay. This is important, because if there are too many false
detections, some will be matched erroneously to real $FUV$
sources. 

In order to match the $FUV$ and $NUV$ catalogues, we first transformed 
the $NUV$ coordinates onto the frame of the $FUV$ mosaic. We then
checked for common pairs within a tolerance of 1.5 pixels; this
resulted in 306 matches. Careful inspection of the registered $NUV$
image around unmatched $FUV$ sources revealed an additional 50 $NUV$
counterparts within this tolerance that had not been detected by 
{\tt daofind}. Increasing the tolerance to 2 pixels added another 23 
matches. These additional matches are located mainly in the outskirts of
the mosaic where our coordinate transformation is slightly less
accurate, thus justifying an increased matching tolerance. All stars
listed in \citet{brown} were found in our $FUV$ catalogue; however, 25
of them did not have $NUV$ matches within 2 pixels. Closer inspection by
eye revealed that these stars are either located in the outer regions 
of the mosaic and had probable $NUV$ counterparts outside the 2 pixel 
tolerance radius, or they had faint $NUV$ counterparts that are
located in the most crowded inner area and were not initially detected
by {\tt daofind}. We therefore added these missing matches by hand.   

A catalogue listing all of our $FUV$ sources is given in
Table~\ref{catalogue}. We include sources without $NUV$ counterparts
in Table~\ref{catalogue}, because these are likely to include
additional WD and CV candidates.\footnote{A $FUV$ source may lack a
  $NUV$ counterpart because (i) it is too faint and blue to show up in
  the $NUV$ mosaic or (ii) a counterpart exists, but was not found due
  to the severe crowding in the $NUV$ image.}
Among the sources with $NUV$ counterparts, we have marked all pairs
that were matched with the higher tolerance of 2 pixel or were added 
by hand. In total, our catalogue contains 521 $FUV$ sources, including 
403 with $NUV$\ counterparts. All 295 sources listed in \citet{brown}
are also contained in our catalogue.

\subsection{Aperture photometry}
\label{photometry}

Aperture photometry was obtained using {\sc daophot}
\citep{stetson}. Following \citet{brown}, we chose an aperture radius
of 4 pixels and used an annulus between 4 and 10 pixel for sky subtraction.  
An exposure map was created that accounts for the
different exposure times at different locations in the composite image. 
For the calibration to the STMAG-system we used the aperture
corrections derived by \citet{brown}, namely 1.83 and PHOTFLAM of
1.11e-16 $\rm{erg}$ $\rm{s}^{-1} \rm{cm}^2 \AA^{-1}/(\rm{counts}$
$\rm{s}^{-1})$ for the F25QTZ data, and 1.44 and PHOTFLAM of 3.92e-17
$\rm{erg}$ $\rm{s}^{-1} \rm{cm}^2 \AA^{-1}/(\rm{counts}$ $\rm{s}^{-1})$
for the fluxes obtained with F25CN270.\footnote{We noticed a typing
  error in \citet{brown}, namely PHOTFLAM in the F25CN270 is 3.92e-17
  $\rm{erg}$ $\rm{s}^{-1} \rm{cm}^2 \AA^{-1}/(\rm{counts}$
  $\rm{s}^{-1})$ (not 3.29e-17 $\rm{erg}$ $\rm{s}^{-1} \rm{cm}^2
  \AA^{-1}/(\rm{counts}$ $\rm{s}^{-1})$)}    

\begin{figure}
\plotone{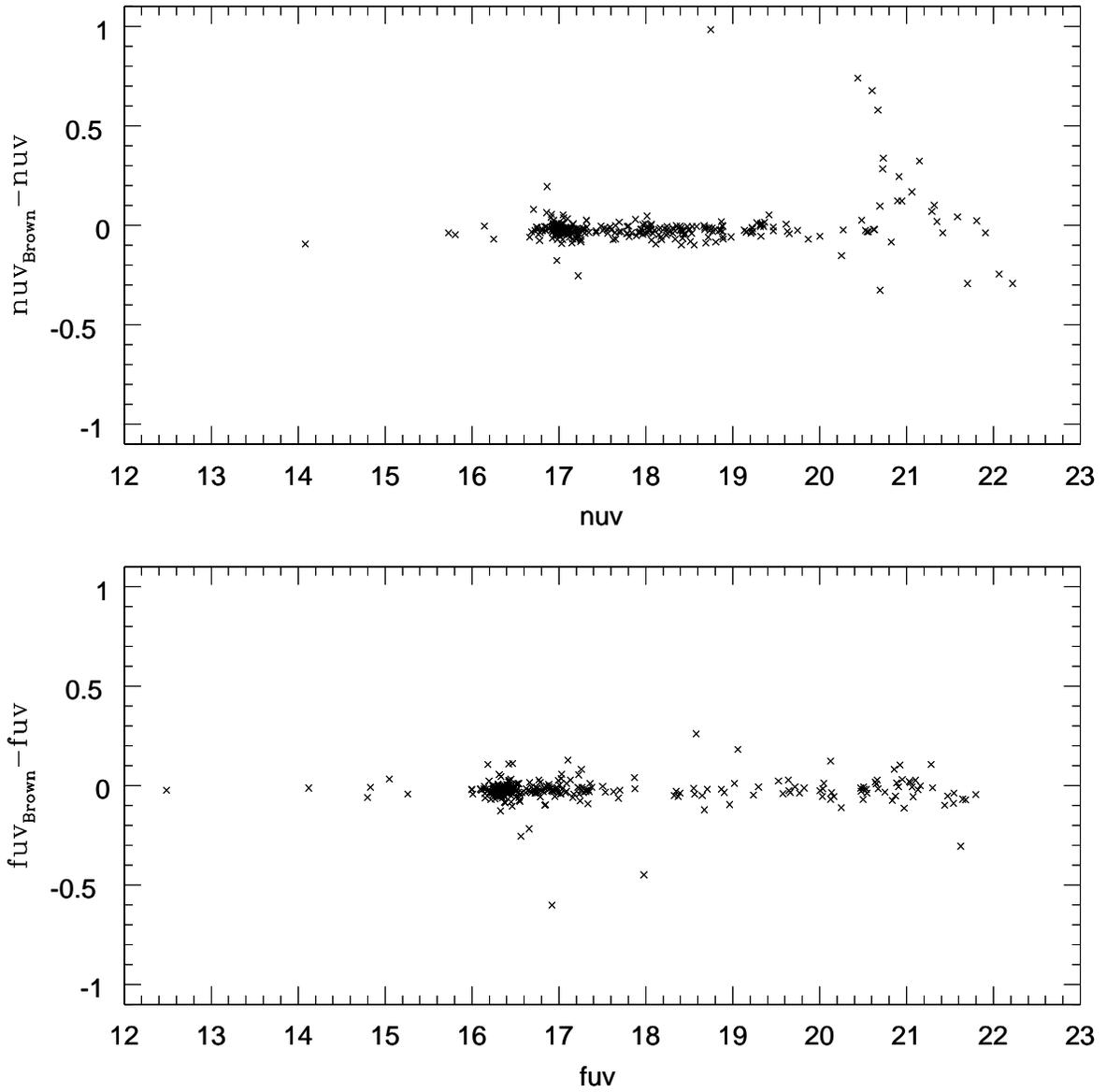}
\caption{\label{centerings} Lower panel: Difference in $FUV$ magnitude
between our photometry and Brown et al.'s (2001). Only a few stars show
differences greater than 0.1 mag, thus in general both datasets show
good agreement. Upper panel: The same but for the $NUV$
magnitudes. See the text for a discussion of outliers.}   
\end{figure}

The results of aperture photometry depend on the coordinates of the
source that has to be measured, or, to be more precise, on the
``centering''. Several re-centering algorithms are available under {\tt
phot} in {\sc daophot} running under {\sc iraf}. \citet{brown} created their
coordinate list by centering on each star by eye. We used their
photometry as a point of reference, and hence adopted 
a Gaussian recentering algorithm, since this yielded the best
agreement with their measurements. Fig.~\ref{centerings} displays the
magnitude differences between our photometry and that of
\citet{brown}, in both $FUV$\ and $NUV$\ bands. Overall, the agreement 
is very good, with a few notable exceptions discussed below.

For star no.\ 516 (Id.\ 29 in Brown et al., 2001) the difference between our
photometry and Brown et al.'s (2001) is 0.6 mag in 
$FUV$. This star is located in the outer regions of the $FUV$\ mosaic
and appears somewhat elongated. This might be because the object is a
blend of two sources, or because the PSF is smeared out since the
coordinate transformations used to construct our mosaic are least
reliable in the outermost regions of the image. The latter effect
would be exacerbated by the asymmetric $FUV$\ PSF and might cause the
brightness to be underestimated when using a small aperture of only 4
pixels. The situation is similar for star no.\ 500 (Id.\ 13 in Brown et
al.\,2001), which is 0.45 mag fainter in our photometry, is also
located close to the edges of the mosaic and appears somewhat
elongated. However, a difference greater than 0.1 mag in the $FUV$
occurred only for 17 stars, and a difference greater than 0.15 mag
only for 5 stars. 

For the $NUV$, we found a magnitude difference of 0.985 mag for star
no.\ 426 (Id.\ 189 in \citet{brown}). This seems to be a case where
\citet{brown} matched a different $NUV$ source than we did. Though we
are quite confident that we correctly matched most of the stars in $FUV$
and $NUV$, this discrepancy is a useful reminder that occasional
mismatches might still occur due to the extreme crowding in the $NUV$
images. This is especially true for fainter sources, and mismatches
probably contribute to the rising $NUV$ differences towards fainter
magnitudes. However, differences greater than 0.15 mag occurred for
only 18 stars in the $NUV$. Setting aside these extreme cases, 
the mean absolute differences between Brown et al.'s (2001) magnitudes
and ours are $0.03 \pm 0.03$ mag in $FUV$ and $0.03 \pm 0.02$ mag in
$NUV$. Thus, in general, there is good agreement between their
catalogue and ours. The remaining differences can be explained by 
the different methods used to construct the mosaic images, to detect
the sources, the different centering algorithms for the photometry and
by the severe crowding in the $NUV$ images. 
 
\section{Results and Discussion}
\label{results}

\subsection{The FUV-NUV CMD}
\label{cmdsection}

\begin{figure}
\plotone{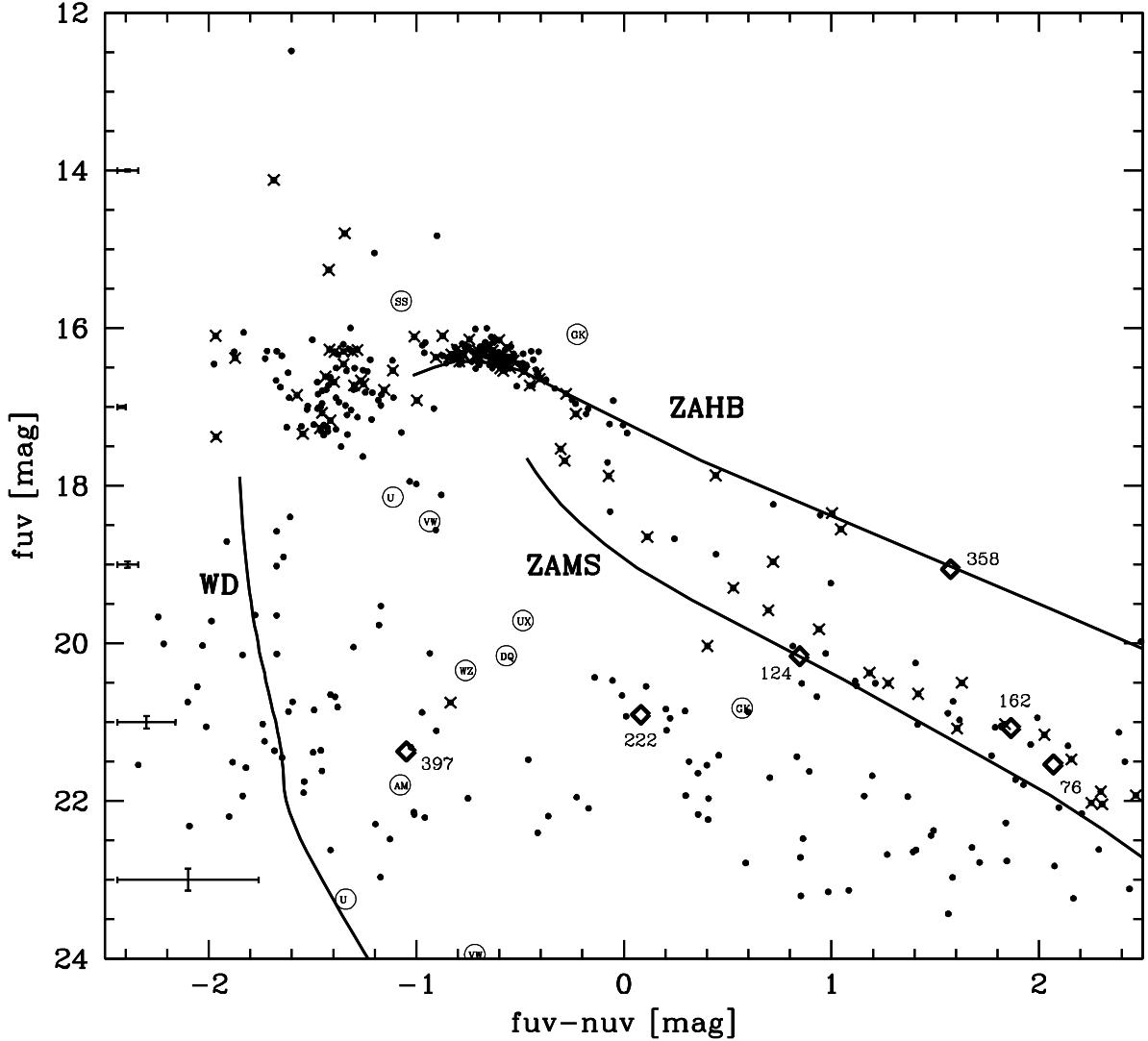}
\caption{\label{cmd}$FUV-NUV$ CMD of the core region of NGC\,2808. 
For orientation purposes, we include a theoretical WD cooling
sequence, a zero-age main sequence, and a zero-age HB track (see text
for details). The diamonds and the corresponding numbers denote
variable $FUV$ sources, as discussed in Sect.~\ref{variability}. 
Mean errors for different $FUV$ magnitude ranges are given on the
left-hand side of the CMD. The open circles with enclosed letters mark
the positions of known field CVs if they were located at the distance
and reddening of NGC\,2808, the notation is the same as in
\citet{knigge1}. The crosses denote stars for which optical
counterparts could be found in the \citet{piotto} data. See the text
for details.}    
\end{figure}

Fig.~\ref{cmd} shows the $FUV-NUV$ CMD of NGC\,2808's core region. 
Several distinct stellar populations can be seen, including blue HB
stars, BSs, WDs, but also CV candidates. Our CMD reaches approximately
2 magnitudes deeper than that presented by \citet{brown}. This is
because we took special care in detecting faint sources in the $FUV$
mosaic (Sect.~\ref{find}). We also note that we plot all sources
independent of their photometric errors, whereas \citet{brown} only
included sources satisfying $m < 22$ mag and $\sigma_m < 0.2$ mag in
both $FUV$ and  $NUV$. However, we stress that only 7 of the sources
plotted have errors exceeding 0.2 mag in $FUV$, and all of these have
errors in the  range 0.2 - 0.3 mag. We therefore still regard these as
significant detections.

To aid in the interpretation of the CMD, we have also calculated and
plotted a set of theoretical tracks. In the following, we will first
briefly describe the construction of the theoretical tracks and then
discuss the various stellar populations within our CMD. 

\subsubsection{Synthetic Photometry}
\label{synthetic}
 
Since main sequence stars and red giants are too cool to show up in our 
CMD, we only present theoretical tracks for the zero-age horizontal
branch (ZAHB), the zero-age main sequence (ZAMS) and the WD cooling
curve in NGC\,2808. 

For our ZAHB track, we interpolated on the grid of oxygen-enhanced
ZAHB models provided by \citet{dorman} to generate a set of models at
the cluster metallicity of $\rm{[Fe/H]} = -1.36$.\footnote{We note
  that the \citet{dorman} grid only extends to the blue HB, not the
  extreme HB (EHB). For a detailed analysis of EHB stars in NGC\,2808,
  the reader is referred to \citet{brown}.} 
We then used {\sc synphot} within {\sc iraf/stsdas} to calculate the
$FUV$ and $NUV$ magnitudes of stars on this sequence. This was 
achieved by interpolating on the Kurucz grid of model stellar
atmospheres and folding the resulting synthetic spectra through the
response of the appropriate filter+detector combinations. 

The ZAMS is included because BSs are expected to lie near and slightly
to the red of this sequence. In order to generate this track, we used
the fitting formulae of \citet{tout} to estimate the appropriate
stellar parameters. The corresponding $FUV$ and $NUV$ colours were
then again estimated from the Kurucz model grid within {\sc synphot}. 

Finally, our theoretical WD sequence was constructed by
interpolating on the \citet{wood} grid of theoretical WD cooling
curves, adopting a mean WD mass of 0.55~M$_{\odot}$. The models were
then again translated to the observational plane by carrying out
synthetic photometry with {\sc synphot}, using a grid of synthetic
DA WD spectra kindly provided by Boris G\"{a}nsicke (see G\"{a}nsicke,
Beuermann \& de Martino 1995). 

For all our synthetic tracks we adopted a distance of 10.2 kpc and a
reddening of $E_{B-V}=0.18\pm0.01$ \citep{bedin}. 

\subsubsection{HB stars}
\label{hb}

NGC\,2808 is one of the most extreme examples among the globular 
clusters with unusual HB morphology, as first noted
by \citet{harris74}. Since then this cluster has received 
considerable attention. 
In general, metallicity is regarded to be the first parameter that
influences HB morphology. However, several globular clusters show
different HB morphologies though their metallicities are similar
\citep{rich}. This led to the suggestion of a second parameter
responsible for the unusual behaviour of the HBs in these 
clusters. Many parameters have been suggested (e.g.\ age, mass loss,
stellar rotation etc.), but a thoroughly convincing explanation is
still lacking 
(see e.g. Rood et al.\ 1993, Recio-Blanco et al.\ 2004). 
NGC\,2808 shows a bimodal HB  and one of the longest blue HB
tails, the so-called extreme HB (EHB), with prominent gaps between the
red HB (RHB), blue HB (BHB) and EHB (e.g. Clement \& Hazen 1989,
Ferraro et al.\ 1990, Byun \& Lee 1991, Sosin et al.\ 1997, Walker
1999, Bedin et al.\ 2000). Only three other globular clusters are known
that show bimodal HBs as well as gaps along their HBs, namely
NGC\,6229, NGC\,6388, and NGC\,6441 \citep{catelan}.  
\citet{sosin} discussed various mechanisms that might be responsible
for the HB multimodality, none of which represents a satisfactory
explanation. They suggested that a combination of effects might result
in such a HB morphology, some of which might be unique to NGC\,2808.    

In our CMD, the two distinct clumps of bright stars around $FUV \sim
16$ mag correspond to the blue ($FUV-NUV \sim -0.7$ mag) 
and extreme ($FUV-NUV \sim -1.2$ mag) HB stars. The red HB stars
are too cool to show up in the $FUV$ or $NUV$ images.
As already reported by \citet{brown}, a gap within the EHB -- as seen in
optical CMDs for this cluster -- is not evident in the $FUV-NUV$
CMD. Instead, the EHB clump is extended towards fainter magnitudes.  
\citet{brown} suggested that these subluminous EHB stars 
(as well as the EHB gap in the optical CMDs) can be explained by a late
helium-core flash that these stars undergo while they descend the
WD sequence. The recent findings of \citet{moehler} and \citet{momany}
generally support this theory. For a comprehensive analysis of the HB
stars in the $FUV-NUV$\ CMD, we refer the reader to \citet{brown}.

\subsubsection{Blue stragglers}

A trail of stars can be seen in Fig~\ref{cmd}, starting from the BHB
clump and reaching down below the theoretical ZAHB sequence towards
the faint red corner of our CMD. These sources show magnitudes fainter
than the ZAHB, but brighter than the ZAMS. The location of these objects
agrees with the expected location of BSs, which are thought to be
dynamically-formed objects resulting from a collision or coalescence
of two or more MS stars. They are more massive than normal cluster MS
stars and therefore we expect them to be already somewhat
evolved. This explains their location slightly above and to the red of
the ZAMS. We find 61 BS candidates, but caution that this should not
be taken as a strict number since the various zones in our CMD partly
overlap, and the discrimination between the CV and BS candidates, in
particular, is difficult (see below).  

\subsubsection{White dwarfs}

Fig.~\ref{cmd} reveals a population of 40 objects that lie
near the theoretical cooling curve and are therefore probably hot,
young WDs. Of these, 22 are brighter than $FUV = 21$ and should be
detectable across the full $FUV$\ and $NUV$\ mosaics. 
How many WDs should we expect to see? Following \citet{knigge1},
we can obtain a simple estimate by scaling from the number of HB stars
in the same field of view. We count $\approx 210$\ BHB and EHB stars in
our CMD, but, as noted above, RHB stars are too cool to show up. We
therefore estimate the total number of HB stars by adopting the ratio
of (RHB+BHB+EHB)/(BHB+EHB)~$= 1.86$\ found by \citet{sosin}. We
therefore expect a total of $\approx 390$ HB stars in our field of
view. Given that the lifetime of stars on the HB is approximately
$\tau_{HB} \simeq 10^{8}$~yrs (e.g. Dorman 1992), we can predict the
number of WDs above a given temperature on the cooling curve from the
relation

$\frac{N_{WD}}{N_{HB}} \sim \frac{\tau_{WD}}{\tau_{HB}}$ 

where $\tau_{WD}$ is the WD cooling age at that temperature (e.g. Knigge
et al.~2002). WDs at $FUV \sim 21$ mag on our WD sequence have 
a temperature of $T_{eff} = 38,000$ K and a corresponding cooling age
of $\tau_{WD} \simeq 4 \times 10^{6}$~yrs. We therefore predict
a population of approximately 16 WDs brighter than $FUV = 21$, which
is in reasonable agreement with the observed number of 22. 
This strongly suggests that most of our candidates are indeed WDs.

\subsubsection{CV candidates}

Fig.~\ref{cmd} reveals quite a number of stars that are located
between the WD and ZAMS tracks. This is a region in which we might
expect to find CVs. We estimate that there are $\simeq 60$ sources in
this ``CV zone''.\footnote{We refrain from giving a more precise
  estimate, since many faint sources have photometric errors that are
  too large to permit a definite assignment to a unique zone. The
  distinction between the red edge of the CV zone and the BSs sequence
  is particularly difficult in this context.} 
How does this number compare to theoretical predictions? We can
attempt to answer this question by scaling the number of tidal capture
CVs predicted for 47\,Tuc to NGC\,2808. To this end, we use 
the simplified estimate for the capture rate in globular cluster cores
\citep{heinke}

$\Gamma \propto \rho_{c}^{1.5} \cdot r_{c}^2 $, 

where $\rho_{c}$ is the central luminosity density and $r_{c}$ the
core radius. Based on this, we find that the total number of
dynamically-formed CVs in NGC\,2808 should be quite similar to 47\,Tuc,
in which \citet{distefano} predicted a population of $\sim 190$ active
CVs formed via tidal capture. For 47\,Tuc, \citet{distefano} found that
approximately half of the captures would take place within 1 core
radius. A CV formed outside the core will drift towards the centre due
to mass segregation. Thus a given active CV should be found in the
core today if its age exceeds the relaxation time-scale at the
location where it formed. Given our detection limits, we cannot hope 
to detect the very oldest CVs, since these are also the faintest. In
order to obtain a rough estimate, we will assume that we can detect
most CVs whose secondaries have not yet been whittled down to brown
dwarfs, but none of the latter. Thus our dividing line is between CVs whose
orbital periods are still decreasing, and the so-called period
bouncers, whose periods are lengthening (see Di\,Stefano \& Rappaport,
1994, for
details). Based on Fig.~2 in \citet{distefano}, the oldest CVs we
can detect will therefore have ages on the order of $10^{9}$~yrs. This
is comparable to the relaxation time-scale at the half-mass radius of
NGC\,2808, so some, but probably not all dynamically-formed active CVs
should have had time to reach the core. Given this ambiguity, we
assume that between 50\% and 100\% of active CVs in NGC\,2808 should
be within our field of view. However, only about half of these are
above our adopted detection limit, yielding a final prediction between
$45 - 95$ CVs. We stress that this is an extremely rough estimate, but it
does bracket the observed number of sources in our CV zone. Thus, a
significant fraction of these candidates might indeed be CVs.

We caution that other sources might also occupy our CV zone.
We have to consider the possibility that mismatches might have
occurred, though special care has been taken in detecting and matching
the $FUV$ and $NUV$ sources (see Sect.~\ref{find} and \ref{photometry}). 
Non-interacting WD/MS binaries without accretion would also
lie in the CV zone. However, according to our ZAMS sequence, a turn-off
star with a mass of 0.85~M$_{\odot}$ would have a $NUV$ magnitude of
$\approx 23$. These objects would therefore be expected in the CV
zone, but close to the WD cooling sequence where we find about 10
objects. If we do not consider these 10 sources, our number of CV
candidates would be rendered to $\approx 50$, which is still
consistent with our above estimates. In principle, our CMD may also
contain background/foreground sources. However, we expect field star
contamination to be negligible (see e.g. Ratnatunga \& Bahcall 1985).

\subsubsection{Cumulative distributions of the stellar populations}

\begin{figure}
\plotone{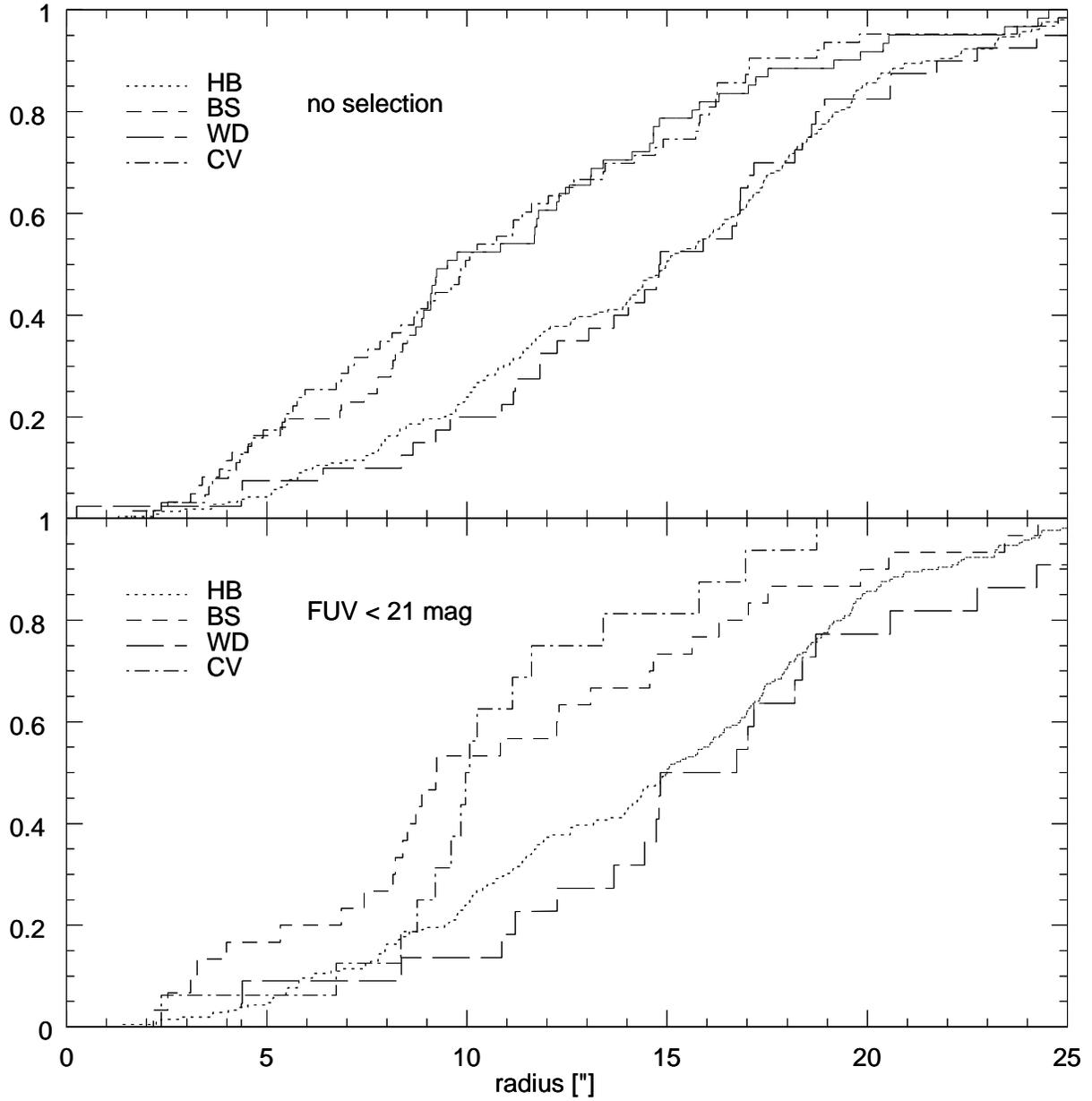}
\caption{\label{cumulative} Cumulative radial distributions of the
  stellar populations that show up in our CMD. Top panel: all sources that
  show up in the corresponding regions in our CMD are considered. Lower
  panel: only sources with $FUV \le 21$ mag are considered. See the
  text for details.}  
\end{figure}

Fig.~\ref{cumulative} shows the cumulative radial distributions of the
stars that show up in our CMD. We distinguish between HB stars (dotted
line), CV candidates (dot - short dashed line), WD candidates (long
dashed line), and BS candidates (short dashed line). The top panel shows the
distribution if {\it all} stars of the corresponding population are
considered, independent of their magnitude. However, as pointed out in
Sect.~\ref{find}, the exposure time is not uniform across our mosaic,
thus also the detection limit varies across the mosaic. For the
outermost regions with the lowest exposure time all sources at least
down to $FUV = 21$ mag are detected. We try to avoid selection effects
by adopting this as a limiting magnitude for our entire dataset. The
lower panel in Fig.~\ref{cumulative} shows the cumulative
distributions of the magnitude-selected data. 

The most striking differences seem to be between our HBs and WD
candidates, which show the least central concentration, and the BSs
and CV candidates, which appear to be the most centrally concentrated
populations. 
This is in agreement with \citet{walker} and \citet{bedin} who
presented deep wide-field photometry and found no radial gradient in
the distribution of the EHB stars.
In order to compare the cumulative distributions and to
establish the statistical significance of the differences between
them, we applied a Kolmogorov-Smirnov test on the magnitude-selected
dataset. However, we caution that the WD, CV and BS candidates cannot
clearly be distinguished on the base of our CMD alone, i.e.\ we might
have sources in one sample that actually belong to another
one. Applying the magnitude selection to our data yields 22 WD, 30 BS,
and 16 CV candidates. The HB stars are brighter than our selection
criterion, and 209 of them are present in our field of view.     

The K-S test returns the probability that the maximum difference between
the two distributions being compared should be as large as observed, under
the null hypothesis that the distributions are drawn from the same parent
populations. According to this, the distributions of HBs and WDs are
consistent with one another, with a K-S probability of 64.7~\%.
However, the distributions of BS and CV candidates differ significantly
from those of the other two populations, with K-S test probabilities of
1.0~\% (CVs vs HBs), 0.5~\% (CVs vs WDs), 0.4~\% (BS vs HBs) and 1.2~\%
(BSs vs WDs). The differences between the CV and BS distributions are
not statistically significant, with a K-S probability of 49.3~\% (CVs vs BSs).

The differences in the observed distributions might reflect the
different masses of the various FUV populations. More specifically, CVs
and BSs are expected to be considerably more massive than WDs and HBs,
since CVs are WD/MS binaries and BSs are roughly speaking MS stars with masses
above the cluster MS turn-off. The heavy CVs and BSs thus sink toward
the cluster core and as a result are more centrally concentrated than
the lighter WDs and HBs. Fig.~\ref{cumulative} shows that the differences
revealed by the K-S test are indeed in the expected direction. 

On the other hand, CVs and BSs are believed to be objects formed
through two- or three-body interactions such as tidal capture,
collisions and mergers. These dynamical interactions take place
preferentially in the dense cluster core. Thus, CVs and BSs may also
be expected to be overabundant in the cluster core because this is
their birthplace.  

In either case, the enhanced central concentration of our CV candidates
provides additional evidence that most of these sources are indeed CVs
or non-interacting WD/MS binaries (as opposed to chance
superpositions, foreground stars, etc), i.e.\ they either segregated
towards the cluster centre, or they formed in the dense core region.   

\subsubsection{FUV sources with optical counterparts}

\citet{piotto} presented optical photometry for 74 Globular clusters,
among them NGC\,2808. The clusters were observed with {\it HST}/WFPC2 in the
$F439W$ and $F555W$ bands, with the PC centred on each cluster's
centre. We searched for optical counterparts to our $FUV$ sources in
their dataset, using {\tt xyxymatch}  within {\sc iraf}. The
transformation between the PC and STIS  physical coordinate
systems is tied to three HB stars as reference objects which could be
clearly identified by eye in both the $F555W$ and the $FUV$
images. Following \citet{knigge1}, we allowed for a maximum difference
of 0.8 PC pixels (corresponding to 1.5 STIS pixels). In total,
we found 97 counterparts to our $FUV$ sources. Out of these
9 are matches with MS stars, red giants or RHB stars. These stars are
too cool to show up in our $FUV-NUV$ CMD, and thus these matches have
to be explained otherwise: 
13156 objects are present in the PC field-of-view
in both optical bands, and the complete PC field is covered by our
$FUV$ mosaic. In addition, 390 $FUV$ sources are located inside the
area covered by the PC-data. Following the same statistical approach as in
\citet{knigge1}, we would expect up to 13 false matches. Thus, all the 
matches with MS stars and red giants can be explained statistically as
false matches. The remaining 88 matches are all blue objects, either
BHB or EHB or BS stars.   

We marked all our $FUV$ sources for which we could find a blue optical
counterpart as crosses in our $FUV-NUV$ CMD, and added their optical
magnitudes and Id from \citet{piotto} to our
Table~\ref{catalogue}. As can be seen, one of our CV candidates, our
star no.~170, does have an optical counterpart with $V = 21.59$ mag
and $B-V = -0.24$ mag. The blue optical colour of the counterpart
(well off the cluster MS), and the quality of the $FUV$/optical match
(the $FUV$ and optical positions agree to within 0.4 pixels) both
suggest that the counterpart identification is secure.

\subsection{Variability}
\label{variability}

As a first check for possible variability among our catalogue stars we
performed photometry on all single $FUV$ images. For this purpose we first
used the processed images that had been used to create the
mosaic, i.e.\ these images were shifted and rotated to a common logical
(i.e. image pixel) coordinate system. The great advantage is that we
can use our coordinate list derived from the $FUV$ mosaic as an input
to all of these processed images. The photometry was carried out in
the same way as described above in Sect.~\ref{photometry}. Of course
only a fraction of our catalogue stars are present in each of the
individual images, and for some stars we could obtain only one measurement. 

\begin{figure}
\plotone{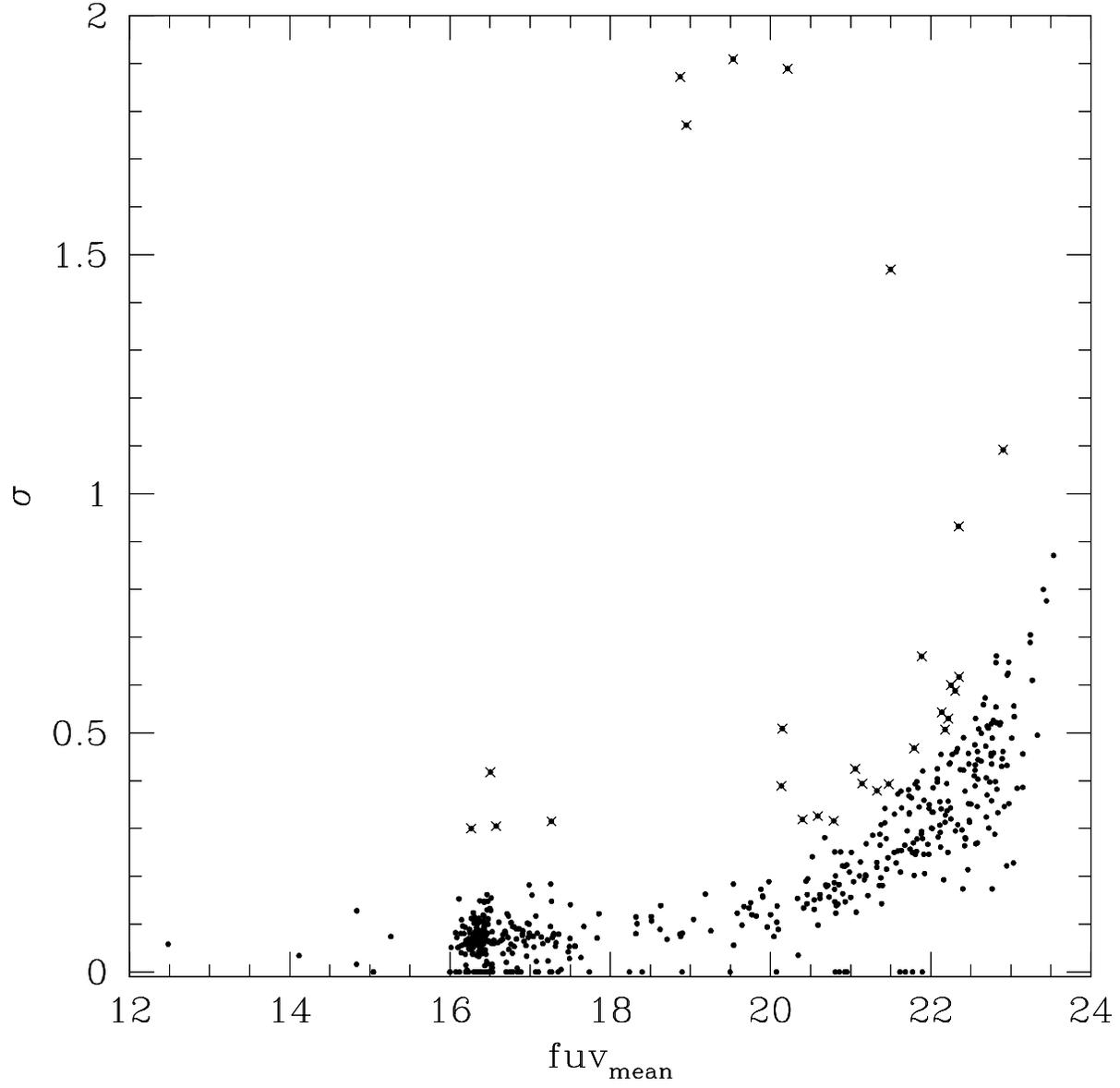}
\caption{\label{varsigma} Mean magnitudes versus
  $\sigma_{mean}$ derived from the individual photometries for each
  star.}  
\end{figure}

Fig.~\ref{varsigma} displays a plot of the mean magnitudes
versus the corresponding $\sigma_{mean}$ calculated from all individual
magnitudes derived for each star. As can be seen, some of our objects
show a considerably higher $\sigma_{mean}$ than the majority. The crosses in
Fig.~\ref{varsigma} denote 28 objects that show a large $\sigma_{mean}$, 
compared to their companions with similar brightness, and that we
chose to inspect more closely. We cross-checked our method using 
median (instead of mean) magnitudes and $\chi^{2}$ (instead of
$\sigma_{mean}$). In each case the same stars show an excess
$\sigma_{mean}$ (or $\chi^{2}$).  

As a next step, we used the non-processed images to get more precise
magnitudes for our 28 variable candidates. This is necessary, because
the orientation of the PSF on these images is the same with respect to
each image and not rotated as for the processed images that were used
for creating the mosaic. Since these images now do not share a
common logical coordinate system, we had to derive the coordinates for
our variable candidates on each single image by using {\tt imexam} under
{\sc iraf}. The photometry was carried out in the same way as
described in Sect.~\ref{photometry}.  

Stars no.~79, 114, 175 and 269 of our Table~\ref{catalogue} show
very high $\sigma_{mean}$, however, these stars are located very close
to bright stars, which makes it difficult to get accurate coordinates
on the single images that have much shorter single exposure times than the
mosaic. Consequently, these stars do not stand out as 
clearly as they do on the $FUV$ mosaic, and sometimes it is not possible
to distinguish them from the halo of the nearby bright star.
For 18 of the other stars that we selected on the basis of
Fig.~\ref{varsigma}, the magnitude scatter either disappears when using the
single non-processed image photometry, or too few measurements are
available for reliable conclusions. However, for none of these 18
sources is a magnitude variation as evident as it is for star no.~358, see
below. 
We note that \citet{proffitt} compared observed and predicted count
rates in the $FUV$ F25QTZ and found a much larger scatter between
observations and predictions than expected from Poisson
statistics. 
Only 6 stars out of the 28 candidates have more than 5
measurements and a $\sigma_{mean}$ well beyond the expected
photometric scatter -- also if we take the effect described by
\citet{proffitt} into account -- and we consider these sources as ``secure''
variable candidates. The remaining 22 stars are labelled in
Table~\ref{catalogue} but are not discussed further here.  

\begin{figure}
\plotone{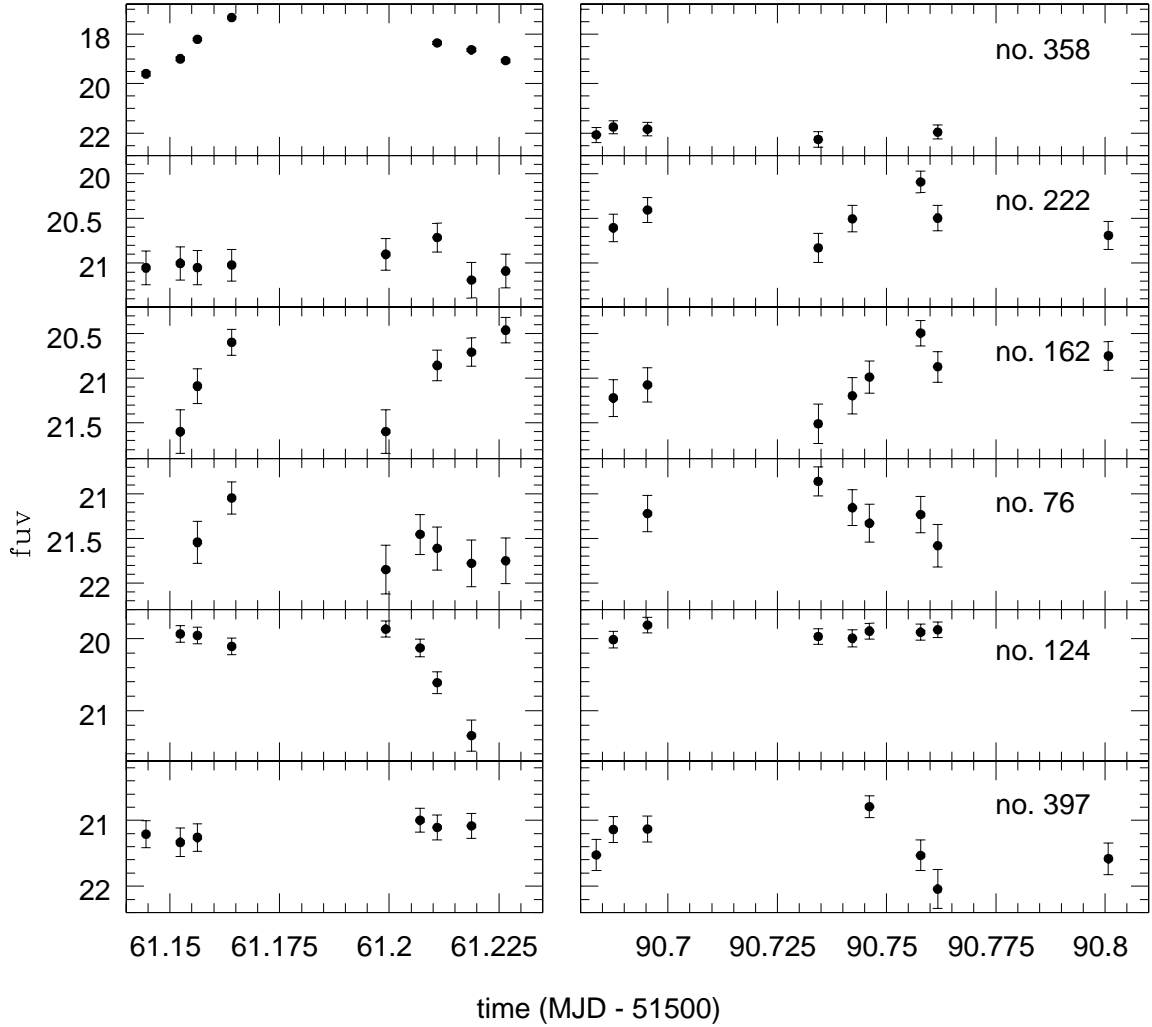}
\caption{\label{varstars} Magnitudes versus time for 6 variable
  candidates. The diagrams have been interrupted for a better display
  of the magnitude variations in each observing period. The top panel
  shows the magnitude-time diagram for star no.~358. This star
  corresponds to the RR Lyrae V\,22 \citep{corwin}. It stands
  out much brighter in the images taken in the first observation interval,
  while it nearly disappears in the second one. See text for more details.
  }  
\end{figure}

NGC\,2808 was observed on only 4 days (18., 19. January 2000 and 16.,
20. February 2000). This period is sufficient to search for signs of
variability, but it is somewhat too short and too interrupted to
present reliable lightcurves for our variable candidates. However,
Fig.~\ref{varstars} shows the magnitudes versus time for the 6 sources
that remain as good variable candidates after our analysis of the
single, non-processed images, i.e.\ these stars still show a
brightness variation that is well outside the expected magnitude scatter. 

The 6 variable candidates are overplotted as diamonds in Fig.~\ref{cmd}. 
Star no.~222 lies in the CV zone at $FUV-NUV = 0.08$ mag, making it a
good CV candidate. Star no.~397 is located close to the WD region in
our CMD, but could certainly be a CV as well. As mentioned above,
the WD and CV zones are likely to overlap and cannot clearly be
distinguished on the base of our CMD alone. 

The lightcurve for star no.~124 in Fig.~\ref{varstars} seems to
suggest that this source might be an eclipsing binary, with the
beginning of an eclipse indicated at the end of the first observing
period. This star is located on our theoretical ZAMS. It is very
likely that this star is a BS star, and some BS are known to be
binaries (see Livio 1993, and references therein). Stars no.~162 and
76 are clearly located in the BS zone of our CMD. Their lightcurves,
especially for star no.~162, seem to indicated that these stars are
pulsating variables (for other examples of oscillating BSs, see
Gilliland et al.\ 1998, and references therein).   

\begin{figure}
\plottwo{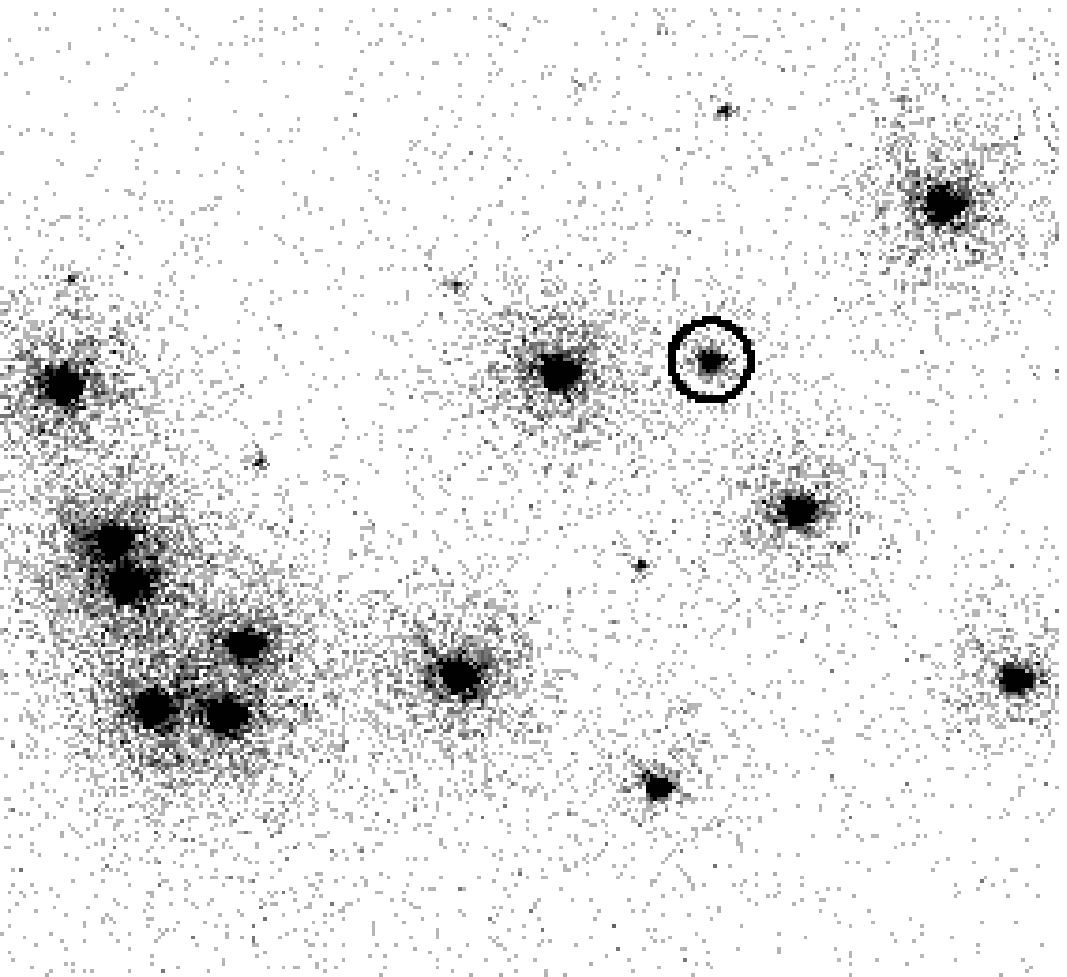}{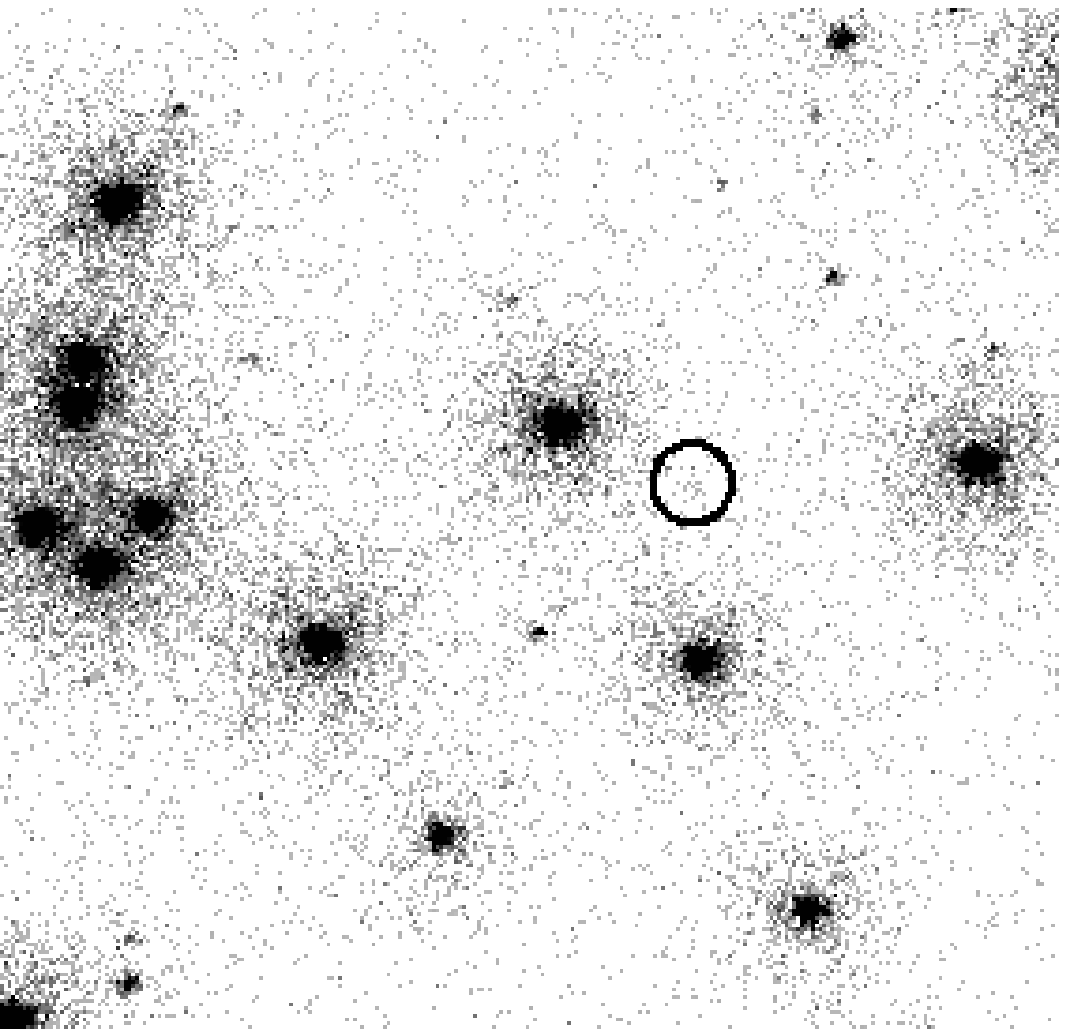}
\caption{\label{no70} Left: Close up of image o60q02f6q centred on
  star no.~358 which clearly sticks out as one of the brighter objects
  ($FUV = 18.212$ mag). Right: Close up of image o60q52kxq, centred
  around the same star, marked with a black circle in both figures. As
  can be seen, star no.~358 has practically vanished.}   
\end{figure}

Star no.~358 is presented in the top panel in Fig.~\ref{varstars}. 
This is the  brightest of our variable candidates, located at $FUV =
19.06$ mag and $FUV-NUV = 1.58$ mag and thus above the BS 
zone. It has a $\sigma_{mean} = 1.9$ mag in Fig.~\ref{varsigma}, and
its brightness drops about 4 magnitudes from the January to the
February observations. This can be seen by eye on the $FUV$ images,
as illustrated in Fig.~\ref{no70}. The image on the left hand-side of
this figure represents a close up of image o60q02f6q. Star no.~358
clearly sticks out as one of the brighter objects ($FUV = 18.212$
mag). The image on the right hand-side of Fig.~\ref{no70} is a
close up of exposure o60q52kxq, centred around the same region. As can
be seen, star no.~358 has practically vanished. We found that the
coordinates of this star agree within $\approx 1 \arcsec$ in $\delta$
with the RR Lyrae V\,22 of \citet{corwin}. \citet{downes} pointed out
that the amplitude of RR Lyrae can be as high as 4 mag in the
$FUV$. Also, the lightcurves of RR Lyrae are known to be asymmetric
with a period of 0.1 to 0.3 days. It seems that we observe a
brightness peak in the first observation period and a minimum in the
second one. Recently, \citet{wheatley} found $FUV$ variations of about
4.9 mag within the 0.56432 d period of their RR Lyrae star, which is of
the same order as the $FUV$ amplitude we found for our star
no.~358. Large $FUV$ amplitudes are to be expected since the
amplitudes of the pulsations increase towards the far-ultraviolet
\citep{downes, wheatley}. The UV photospheric flux is extremely
sensitive to $T_{eff}$ in the temperature range of RR Lyrae stars
($T_{eff} \sim 6500$ K).  
 
\section{Summary}
\label{summary}

We have reanalyzed far-UV STIS {\it HST} observations of the core
region of the globular cluster NGC\,2808. These data were first
analyzed by \citet{brown} with an emphasis on the bright BHB and EHB
stars, whereas our focus has been on the population of fainter $FUV$
sources. Taking special care in detecting the faint $FUV$ sources, we
were able to present a $FUV-NUV$ CMD that is $\approx 2$ mag deeper
than the one presented by \citet{brown}. The overall agreement between
their photometry and ours is good. 

Various stellar populations show up in our CMD, including the BHB and EHB
stars for which this cluster is well known. RHB and MS stars are too
cool to show up. However, approximately 60 BS candidates are present
in our CMD. About 40 sources are located close to our theoretical WD cooling
curve and are probably hot, young WDs. We estimated the number of WDs
by scaling from the number of HB stars that are present in our
field of view, and found that approximately 16 WDs can be expected at
$FUV \le 21$ mag. This is in reasonable agreement with the observed
number of 22 at a completeness limit of 21 mag. 
Our CMD reveals a number of stars that are located in the CV region
between the WD cooling curve and the ZAMS. Approximately 60 sources
can be found in this region, but we caution that due to the
photometric scatter the discrimination between the individual zones
can be difficult. Using a simplified estimate for the capture rate in
globular cluster cores \citep{heinke}, we found that the number of
dynamically-formed CVs in the core of NGC\,2808 is comparable to
47\,Tuc, and broadly consistent with the number of candidates we
detected. This suggests that many of our candidates might indeed be CVs.
The cumulative radial distributions of the individual stellar
populations indicate that our BS and CV candidates are more centrally
concentrated than HBs and the WD candidates. This may be a result of
mass segregation, but might also reflect the fact that CVs and BS are
mostly dynamically-formed populations that were preferentially born in
the dense cluster core. 
We compared our $FUV$ data with optical {\it HST}/WFPC2 data presented by
\citet{piotto} and found 88 optical counterparts for our HB and BS
stars and for one CV candidate (our star no.~170).

We searched for variability among our catalogue stars and found six
good variable candidates. These stars are all located at $FUV \ge 19$
mag and distributed over the entire colour range of our CMD. Two
variable candidates, stars no.~222 and 397, lie in the CV zone of our
CMD and thus are excellent CV candidates. Stars no.~124, 162 and 76 are all
located in the BS zone of our CMD. The lightcurve of star no.~124
suggests that this source may be an eclipsing binary, while the other two
stars might be pulsating variables.

The most striking among our variable stars is our star no.~358 . This
is the brightest variable in our sample and shows a considerable drop
off about 4 mag during the observation period. We identified this star
as the RR Lyrae V\,22 in \citet{corwin}. RR Lyrae can show large $FUV$
amplitude variations of even up to $6 - 8$ mag (see Downes et al.~2004,
Wheatley et al.~2004).

Overall, the results of our study confirm that $FUV$ observations are a
powerful tool for studying hot, and especially dynamically-formed, stellar
populations in the cores of GCs.

\acknowledgments
We are grateful to an anonymous referee for a detailed report that
helped us to improve this paper.

\newpage

\setlength{\tabcolsep}{0.11cm}
\begin{table}
\begin{center}
\caption{\label{catalogue} Catalogue of all FUV-NUV sources. $\delta
  FUV$ and $\delta NUV$ denote the difference between our $FUV$, $NUV$
  photometry and the one presented by \citet{brown}. Our $\alpha$
  $\delta$ coordinates are based on the coordinate system given by the
  FITS-headers, but have been shifted to achieve the best match to the
  \citet{brown} coordinates. This required offsets of $-0\farcs13$ in
  $\alpha$ and $-0\farcs6$ in $\delta$, after which the rms differences
  between our coordinates and Brown et al.'s~(2001) are only
  $0\farcs006$ in $\alpha$ and $0\farcs048$ in  $\delta$. However, as
  noted in \citet{brown}, although the relative astrometry is very
  accurate, the error in the {\em absolute} astrometry is much larger
  ($\approx 1\arcsec - 2 \arcsec$). 
  Column 10 of our table lists the distance to the cluster centre
  which we determined from star number counts to be at 
  $9^{\rm{h}}$~$12^{\rm m}$~$2.96^{\rm s}$ and 
  $-64^{\circ}$~$51\arcmin$~$47.78\arcsec$. 
  Only the first 20 sources are listed, the complete
  catalogue is available at CDS, Strasbourg.} 
\begin{tiny}
\begin{tabular}{rcccccccccccccl}
\tableline
no.  & alpha & delta & $FUV$ & $\Delta FUV$ & $NUV$ & $\Delta NUV$ &
$\delta FUV$ & $\delta NUV$ & radius & $\rm{Id}_{Brown}$ & $\rm{Id}_{Piotto}$ & $V$ & $B$ & comment \\ 
 & [$^{\rm{h}} ~^{\rm m} ~^{\rm s}$] & [$^{\circ} ~\arcmin ~\arcsec$] & [mag]  & [mag] & [mag] & [mag] & [mag] & [mag] & [$\arcsec$] & & & [mag] & [mag] & \\\hline
 1 & 9 12 5.454 & -64 51 26.761 & 20.808 & 0.170 & 22.187 & 0.229 &      - &      - & 26.51 &   - & - & - & - &  a   \\       
 2 & 9 12 5.313 & -64 51 26.404 & 20.074 & 0.121 &      - &     - &      - &      - & 26.28 &   - & - & - & - &      \\     
 3 & 9 12 4.414 & -64 51 25.250 & 21.729 & 0.260 & 19.842 & 0.062 &      - &      - & 24.54 &   - & - & - & - &      \\     
 4 & 9 12 5.149 & -64 51 28.025 & 17.131 & 0.023 & 18.415 & 0.022 &  0.029 & -0.025 & 24.34 & 253 & - & - & - &  pv  \\     
 5 & 9 12 5.969 & -64 51 30.908 & 22.210 & 0.229 & 23.169 & 0.379 &      - &      - & 25.69 &   - & - & - & - &  pv  \\     
 6 & 9 12 4.898 & -64 51 27.503 & 21.911 & 0.198 &      - &     - &      - &      - & 23.92 &   - & - & - & - &      \\     
 7 & 9 12 6.306 & -64 51 32.282 & 18.579 & 0.044 & 20.252 & 0.073 &  0.261 & -0.152 & 26.50 & 289 & - & - & - &  n   \\     
 8 & 9 12 5.210 & -64 51 28.903 & 16.468 & 0.017 & 16.938 & 0.011 &  0.112 &  0.022 & 23.88 & 256 & - & - & - &  pv, e\\   
 9 & 9 12 6.131 & -64 51 32.254 & 16.448 & 0.017 & 17.148 & 0.012 &  0.032 & -0.088 & 25.63 & 287 & - & - & - &  e   \\      
10 & 9 12 5.059 & -64 51 28.739 & 18.869 & 0.050 & 18.425 & 0.022 & -0.019 & -0.065 & 23.43 & 252 & - & - & - &      \\     
11 & 9 12 4.342 & -64 51 26.459 & 16.422 & 0.023 & 17.003 & 0.016 &  0.108 & -0.033 & 23.26 & 219 & - & - & - &  n, e\\     
12 & 9 12 6.749 & -64 51 34.589 & 16.458 & 0.023 &      - &     - &      - &      - & 27.63 &   - & - & - & - &  e   \\      
13 & 9 12 4.119 & -64 51 25.910 & 21.532 & 0.275 &      - &     - &      - &      - & 23.26 &   - & - & - & - &      \\     
14 & 9 12 5.746 & -64 51 31.458 & 20.506 & 0.105 & 19.294 & 0.033 &      - &      - & 24.27 &   - & - & - & - &  pv  \\      
15 & 9 12 5.530 & -64 51 30.936 & 19.978 & 0.082 & 17.486 & 0.014 &      - &      - & 23.65 &   - & - & - & - &  pv  \\     
16 & 9 12 3.889 & -64 51 25.800 & 19.452 & 0.091 &      - &     - &      - &      - & 22.96 &   - & - & - & - &      \\    
17 & 9 12 3.460 & -64 51 24.399 & 16.507 & 0.017 & 17.141 & 0.012 & -0.057 & -0.041 & 23.80 & 153 & - & - & - &  n, e\\    
18 & 9 12 3.147 & -64 51 23.712 & 16.475 & 0.023 & 17.072 & 0.016 & -0.005 & -0.032 & 24.31 & 128 & - & - & - &  a   \\    
19 & 9 12 3.336 & -64 51 24.893 & 16.422 & 0.016 & 17.201 & 0.012 &  0.018 & -0.071 & 23.20 & 145 & - & - & - &  n, e\\      
20 & 9 12 5.822 & -64 51 33.490 & 16.325 & 0.016 & 16.810 & 0.010 & -0.025 & -0.020 & 23.29 & 280 & - & - & - &      \\    
\tableline
\end{tabular}
\end{tiny}
\normalsize
\tablenotetext{}{a: The tolerance radius for matching these sources in FUV and
  NUV was increased from 1.5 to 2 pixel.}
\tablenotetext{}{n: These pairs were added by hand, see text for
  details.}
\tablenotetext{}{e: These stars show a somewhat elongated shape on the
  $FUV$ mosaic and are located mainly in its outer regions.}
\tablenotetext{}{v: Good variable candidate, see text for details.}
\tablenotetext{}{pv: Possible variable candidate, but too few measurements
  for a reliable statement.}
\tablenotetext{}{nv: Unsure variable candidate, measurements on the
  single images agree within the errors.}
\tablenotetext{}{pv-f: Unsure variable candidate: faint companion to a
  bright star. Measurements on the single images are extremely
  difficult and thus not very reliable.}
\end{center}
\end{table}

\end{document}